 \documentclass[twocolumn]{aastex6}

\usepackage{amsmath}
\usepackage{graphicx}
\usepackage{rotating}
\usepackage{natbib}

\def\be{\begin{equation}}
\def\ee{\end{equation}}  
\def\ba{\begin{eqnarray}}
\def\ea{\end{eqnarray}}  
\def\ro{R_{\rm out}}

\newcommand{\bs}[1]{\boldsymbol{#1}}
\newcommand{\BB}{\bs{\mathcal{B}}}

\newcommand{\BT}{\bs{B}_{\rm T}}

\newcommand{\dome}{d\Omega/\Omega_{eq}}
\newcommand{\comega}{C_\Omega }
\newcommand{\calpha}{C_\alpha}
\newcommand{\cur}{C_u}
\newcommand{\clambda}{C_\omega}
\newcommand{\rat}{E_{\rm T}/E_{\rm tot}}

\def\Eq#1{Eq.~(\ref{#1})}
\def\fig#1{fig.~(\ref{#1})}

\def\s0#1#2{\mbox{\small{$ \frac{#1}{#2} $}}}
\def\0#1#2{\frac{#1}{#2}}

\begin{document}

\title{Stellar dynamo models with prominent surface toroidal fields}

\author{Alfio Bonanno}
\affil{INAF - Osservatorio Astrofisico di Catania, via S.Sofia 78, 95123, Catania, Italy}

\begin{abstract}
Recent spectro-polarimetric observations of solar-type stars have  
shown the presence of photospheric magnetic fields with a predominant toroidal component. 
If the external field is assumed to be current-free it is impossible to explain these 
observations within the framework of standard mean-field dynamo theory.
In this work it will be shown that if the coronal field of these stars is assumed to be harmonic, 
the underlying stellar dynamo mechanism can support photospheric magnetic fields
with a prominent toroidal component even in the presence  of axisymmetric 
magnetic topologies. In particular it is argued that the observed increase 
in the toroidal energy in low mass fast rotating stars can be naturally explained with an underlying $\alpha\Omega$ mechanism.
\end{abstract}

\keywords{dynamo - magnetohydrodynamics (MHD) - stars: magnetic fields - stars: coronae}

\section{Introduction}
One of  the most compelling problems in modern dynamo theory is 
the  formulation of a realistic  coupling between the internal magnetic field and the external field in the atmosphere.
In fact the boundary conditions for the electric and the magnetic field 
at the stellar surface  put  severe constraints on the allowed coronal field configurations,
and it is often necessary to resort to very crude approximations for the latter. 

The standard textbook boundary condition employed in mean-field dynamo theory amounts to consider a current-free field in the region
$r\geq R$ where $R$ is the stellar radius, so that $\nabla\times\mathbf{B}=0$ in this domain. 
Although in the solar case this assumption is motivated  by the possibility of describing the almost rigid 
rotation of the coronal holes in the lower corona \citep{nash88}, it be might incorrect to extend its validity in more active stars.

In fact, as current-free fields represents the states of minimum energy under the constraint that the normal component
of the field at the photosphere is fixed,  they cannot provide the additional energy required to
sustain a significant activity level. Recent measurements of Faraday rotation in the solar corona
support the evidence for large scale coronal currents \citep{spa08}, an essential ingredient to explain coronal heating
in terms of Joule dissipation.  On the other hand on much smaller scales the presence of currents is unavoidable 
in order to explain the twisted field structure of filaments and prominences.

Force-free magnetic fields, defined by $\nabla\times\mathbf{B}=\alpha_{ff}(\mathbf{x})\mathbf{B}$ where $\alpha_{ff}(\mathbf{x})$ is a scalar function, 
can be more appealing from the physical point of view, at least for very low plasma-$\beta$ values.
However, recent investigations based on direct numerical simulations have shown that the free magnetic energy and the efficiency of coronal heating 
via currents dissipation are still very limited in these models \citep{warne}. 

An important consequence of  current-free boundary conditions is that the toroidal field must 
identically vanish in the all domain  $r\geq R$.  In order to illustrate this point in detail it is convenient to introduce 
the scalar  potential $\Phi(r,\vartheta,\varphi)$  so that the toroidal field can be
written as 
\be
\BT= -\frac{1}{\sin \theta}\frac{\partial\Phi}{\partial \varphi}\mathbf{e}_\vartheta+\frac{\partial \Phi}{\partial\vartheta}\mathbf{e}_\varphi . 
\ee
It is not difficult to show that, if $\Phi$ is a single valued  function, 
$\nabla\times \BT = 0$ in a volume always necessarily implies $\BT\equiv 0$ everywhere \citep{krause80}.
In particular, if the field configuration is spherically symmetric the azimuthal component of the magnetic field must 
vanish at the surface,
so that $\frac{\partial \Phi}{\partial\vartheta}|_{r=R}=0$, 
as imposed in most of the dynamo models (see \cite{moss09} for an interesting discussion on this issue). 

Spectropolarimetric observations of photospheric magnetic fields in solar-like stars have  revealed surface toroidal field which 
are mostly axysimmetric and have a predominant toroidal component \citep{petit05,petit08,fares10,fares13,see15}
implying  that most of the magnetic energy resides in the toroidal field. 
This is the case of the solar-like stars like  HD72905, with 82\% of the magnetic energy stored in the toroidal field which is 
nearly completely axisymmetric (97\%), or of the G8 dwarf 
$\xi$ Boo A with 81\% of toroidal energy of which 97\% is due to the axisymmetric component, 
or the case of HD56124 with 90\% of the energy in axisymmetric field configurations, and roughly the same strength of poloidal and 
toroidal component. The situation is even more dramatic if one considers M dwarfs like WX UMa or AD Leo where
nearly all the energy is stored in an axysymmetric field with prominent photospheric non-zero toroidal component (see \cite{see15} for a 
detailed discussion). 

While in the case of the Sun a similar observational strategy has confirmed that the magnetic energy is mostly ($>90$\%) poloidal
\citep{vidotto16}, it is clear that the boundary conditions based on current-free coronal field might not be correct  
for other, more active, stars.  In fact 
it has been further  noticed  that fast rotating, low mass stars have on  average stronger surface toroidal fields
than solar-mass slow rotators \citep{see16}.
Can this fact be explained as an enhanced dynamo action due a  $\alpha\Omega$ mechanism?
Indeed, as strong toroidal fields can alter the average atmospheric structure there is no reason
to assume that a current-free field is still a reasonable approximation to discuss
the magnetic energy budget in these objects. 

The idea proposed in this paper is that
on large scales  and on time scales of the order of  the stellar cycle, 
the external field can be considered harmonic so that:
\be\label{h1}
\nabla^2 \mathbf{B}+k^2 \mathbf{B}=0 
\ee
{where the wavenumber $k$, assumed to be real, determines the characteristic spatial length of the field.

As it is well known, in the Sun various MHD instabilities trigger multiple modes harmonics on different length scales \citep{naka05,arre13}. 
In more active stars one can thus argue that the dynamo waves from the interior trigger global 
oscillations of the coronal field which then propagate according to \eqref{h1}. 

It is important to remark that linear force-free fields with $\alpha_{ff}^2=k^2$
are  solutions of \eqref{h1} 
(notice that strictly speaking $\alpha_{ff}$ is a pseudoscalar  while $k$ is a scalar), 
although the converse is not true in general \citep{chandra57}. In this investigation 
it turns out  that the  coronal fields obtained by coupling \eqref{h1} with  dynamo solutions in the interior
are approximately force-free in the sense of \cite{warne10} as  
$\langle (\mathbf{J}\times \mathbf{B})^2 \rangle \ll \langle \mathbf{B}^2\rangle \langle \mathbf{J}^2 \rangle$.}
\footnote{Notice that imposing  the external field to be force-free 
requires a vanishing toroidal field at the surface for mathematical compatibility with the interior solution, 
as discussed in \cite{reye99} for an $\alpha\Omega$ dynamo.}

It will then be shown that standard MHD continuity conditions at the surface naturally allow for 
non-zero surface toroidal fields which is directly extrapolated from the interior field produced by the dynamo. 
In this letter several  solutions obtained  for solar-like stars will therefore be discussed and analyzed. 
It turns out that depending on the strength of the differential rotation and the surface meridional circulation,  
it is possible to obtain surface fields whose energies distribution between toroidal 
and poloidal components is consistent with the observations. 
 
\begin{figure*}
\begin{center}
 \includegraphics[width=0.32\textwidth, angle=0]{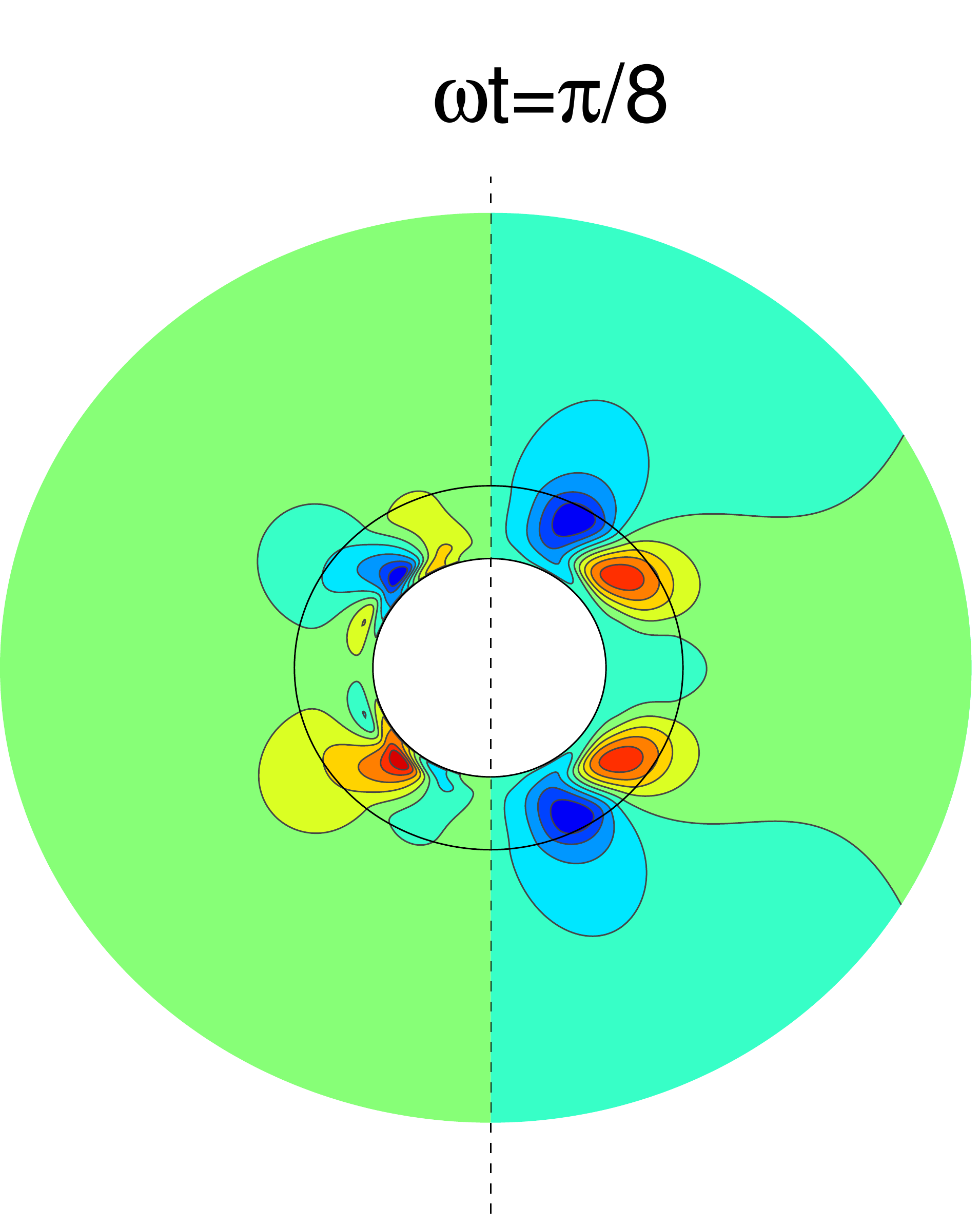}
 \includegraphics[width=0.32\textwidth, angle=0]{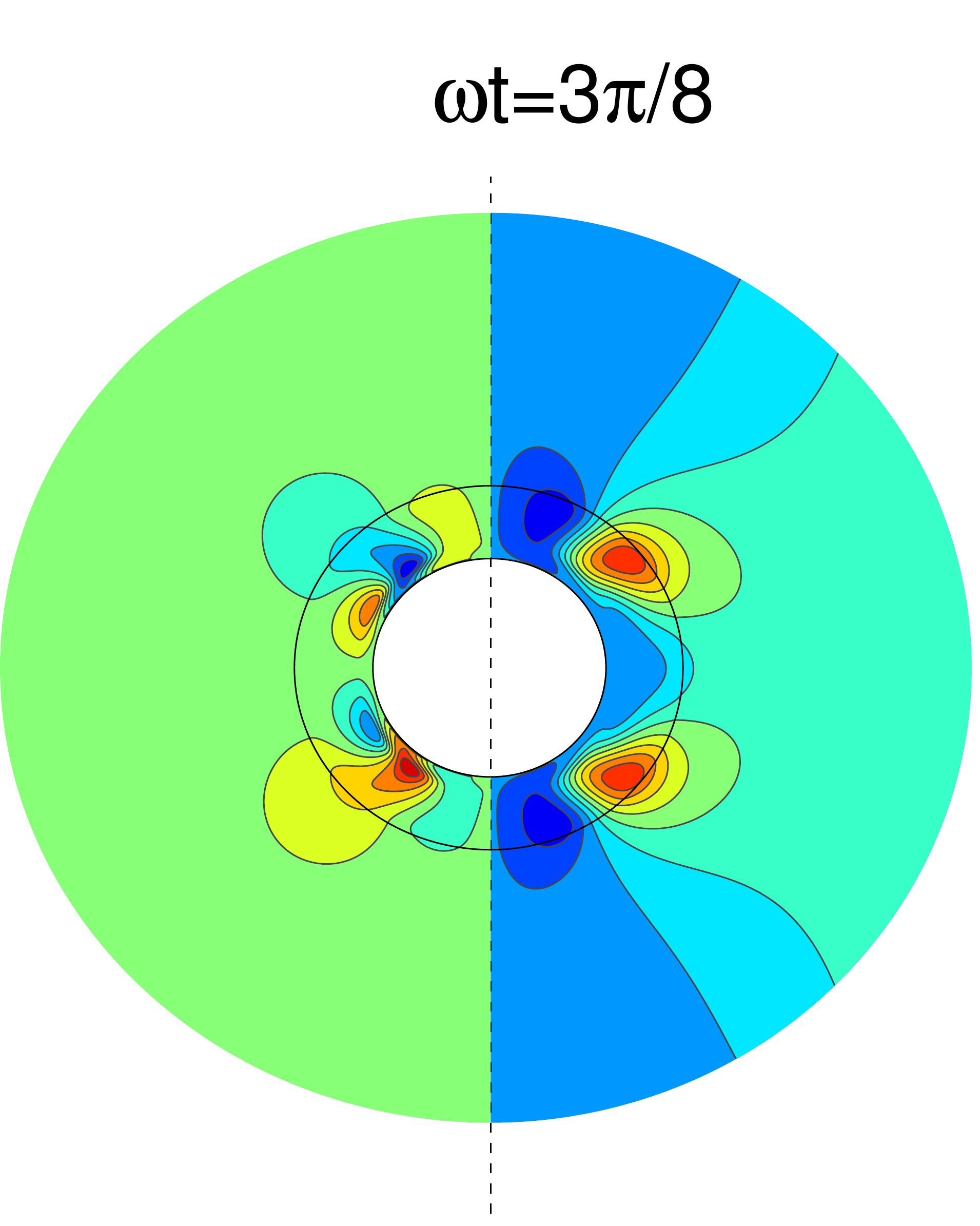}
 \includegraphics[width=0.32\textwidth, angle=0]{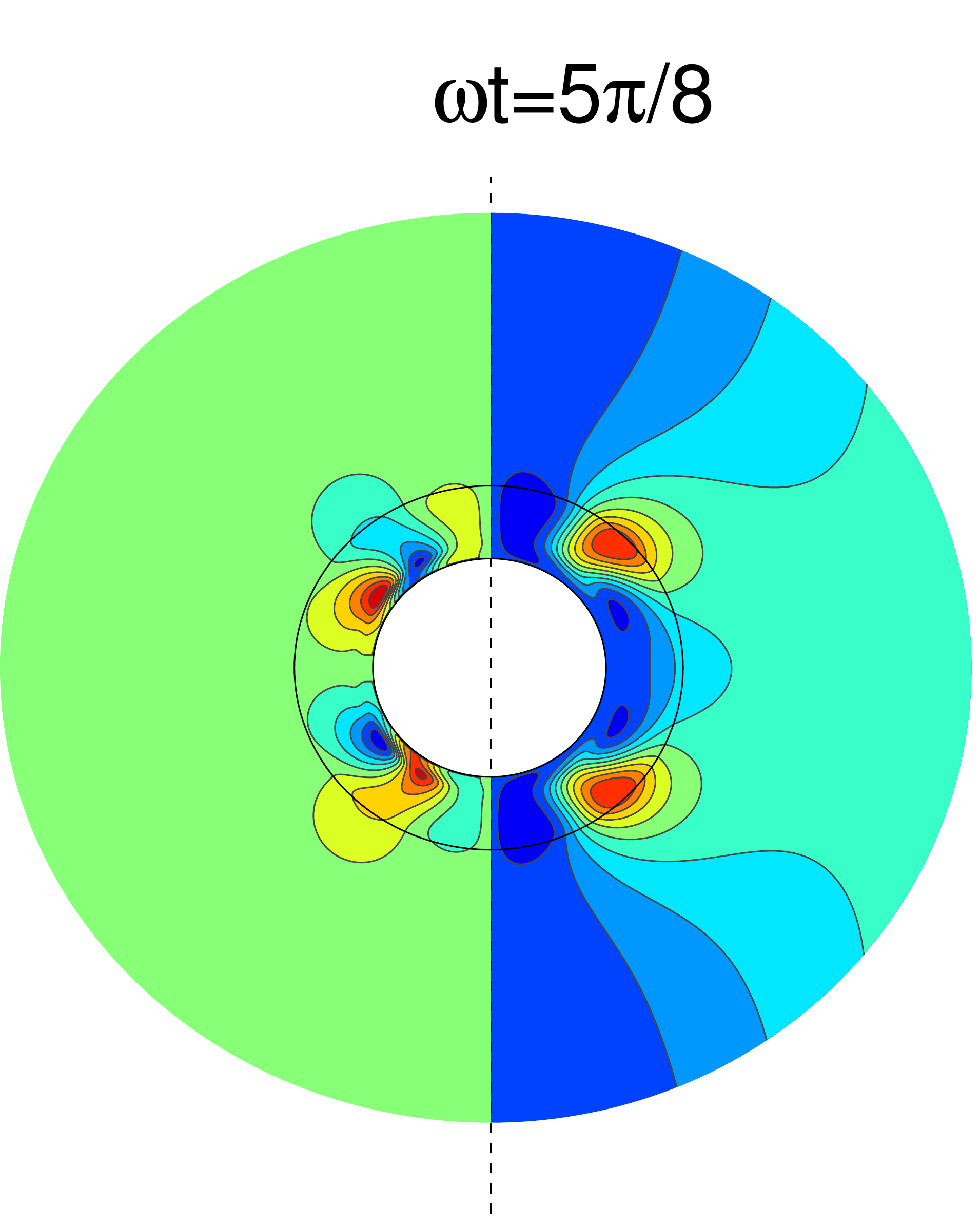}
\caption{Temporal evolution of the global (interior+corona) solution for model J of table (\ref{tab1}). 
The left hemisphere represents the isocontour lines of the toroidal field with blue levels 
for negative $B_\varphi$ and red for positive values of the field. The right hemisphere represents the 
streamlines of the poloidal field. Blue levels are for counterclockwise field lines, red levels for clockwise field lines. 
Notice the opening of the field lines at $2R$. 
\label{fig1}}
\end{center}
\end{figure*}
\section{Basic equations}
Let us assume that  the field periodically  evolves with a characteristic cycle frequency $\omega$  so that  
$\mathbf{B}=e^{-i \omega t} \BB$.    \Eq{h1} thus reads
\be\label{helm}
(\nabla^2 + k^2)\BB =0
\ee
where the wavenumber $k$ is considered to be real. 
Clearly we must assume that $R k\ll 1$ otherwise the typical spatial structure of the field would be too short 
to be consistent with our quasi-homogeneous  approximation. As we shall see, as long as $R k \ll 1$ 
our results are not quantitatively dependent on the value of $k$.

As usual, the boundary conditions are  the continuity 
of the normal component of the magnetic field and the tangential component of the electric field across the stellar surface, 
\be\label{boundary}
[[\mathbf{n} \cdot \mathbf{B}]]=0,\;\; [[\mathbf{n} \times \mathbf{E}]]=0
\ee
where $\mathbf{n}$ is the normal to the surface. 
In spherical symmetry the following decomposition for the magnetic field $\BB$ can be used
\be\label{deco}
\BB=-\mathbf{r} \times \nabla \Psi - \nabla\times (\mathbf{r}\times \nabla \Phi)\equiv
\BB_{\rm T}+\BB_{\rm P}
\ee
where $\Psi=\Psi(r,\vartheta,\varphi)$ and 
$\Phi=\Phi(r,\vartheta, \varphi)$ are scalar functions,  $\BB_{\rm T}$ is the toroidal component
and $\BB_{\rm P}$ is the poloidal one (see \cite{krause80} for details). 
The vector Helmholtz equation \eqref{helm} decouples in the two scalar equations
\begin{subequations}
\label{hel}
\ba\label{hela}
&&\nabla^2 \Phi+k^2\Phi=0\\
&&\nabla^2 \Psi+k^2\Psi=0
\label{helb}
\ea
\end{subequations}
and variable separation in \eqref{hel} gives
\begin{subequations}
\label{d1}
\ba
&&\Phi=R \sum_{n}^\infty[A_{n}j_n{(\xi x)}+B_{n}y_n(\xi x)]P_n(\cos\theta) \\
&&\Psi=\sum_{n}^\infty[C_{n}j_n{(\xi x)}+D_{n}y_n(\xi x)]P_n(\cos\theta)
\ea
\end{subequations}
where $x$ is the normalized stellar radius $x=r/R$, $\xi=k R$,  
$P_m(\cos\theta) $ are the Legendre polynomials  and 
$j_n(x)$ and $y_n(x)$ are the spherical Bessel functions.
{It should be remarked that $k=0$  in \eqref{hel} does not necessarily imply 
$\Psi\equiv 0$ in the all $r\geq R$ domain, as it must instead hold in the case of a current-free  field
(vacuum boundary condition)}. 

The $A_n$-$D_n$ constants are complex numbers  whose value must be determined 
by the boundary conditions \eqref{boundary} imposed at the stellar surface and
at some finite outer radius $r=\ro$ where a transition to a wind dominated 
field topology occur \citep{parker58}.
For our purposes it will be sufficient to assume that 
at $r=\ro$ the solution is radially dominated so that\footnote{Notice that in principle it would be possible to explicitly extend our solution
beyond $r=\ro$ by matching it with  Parker's wind solution, so that the field has the expected $1/r^2$ decay at large distances.} 
\be
\label{obound}
B_\theta(r=\ro)=B_\phi(r=\ro)=0.
\ee
In  the stellar interior the field is described by the mean-field dynamo  equation 
\be\label{induction}
\frac{\partial\mathbf{B}}{\partial t} = \nabla\times(\mathbf{U}\times\mathbf{B}+\alpha \mathbf{B})-
\nabla\times(\eta\nabla\times\mathbf{B})
\ee
where, as usual, $\alpha$ is a pseudo-scalar function representing the turbulent $\alpha$-effect, 
$\mathbf{U}$ is the mean flow and $\eta$ is the turbulent (eddy) diffusivity. 
As before  $\mathbf{B}=e^{-i\omega t}\BB$
and we apply the fundamental decomposition \eqref{deco} for $\BB$.
At last we write
\begin{subequations}
\label{d2}
\ba
\Phi=R \sum_{n}^\infty\phi_{n}(x)P_n(\cos\theta) &&\\
\Psi=\sum_{n}^\infty\psi_{n}(x)P_n(\cos\theta)  && 
\ea
\end{subequations}
where $\psi_{n}$ and $\phi_{n}$ are the complex eigenfunctions of antisymmetric parity of the linear operator  \eqref{induction}
(see  \cite{rad73} for details). 

The field at the inner boundary is assumed to be a perfect conductor, but the 
boundary conditions \eqref{boundary} at the surface imply  
the continuity of  $\phi_{n}$, $\psi_{n}$ and their derivatives across the boundary. 
It is not difficult to show that, 
in order for the interior solution to be consistent with the external field in  \eqref{d1}
the following relation must hold at $x=1$ :
\be
\label{boundary2}
\frac{d \phi_{n}}{dx} +\left [\frac{\gamma_{n} y_{n+3/2}(\xi)+j_{n+3/2}(\xi)}{\gamma_{n} y_{n+1/2}(\xi)+j_{n+1/2}(\xi)}-n \right]\phi_{n}  = 0
\ee
where $\gamma_{n}=A_n/B_n$ are determined by imposing the outer boundary condition \eqref{obound} 
on $r=\ro$.  A similar equation can be obtained for the $\psi_{n}$ components. 
\begin{figure}
\begin{center}
 \includegraphics[width=0.32\textwidth, angle=0]{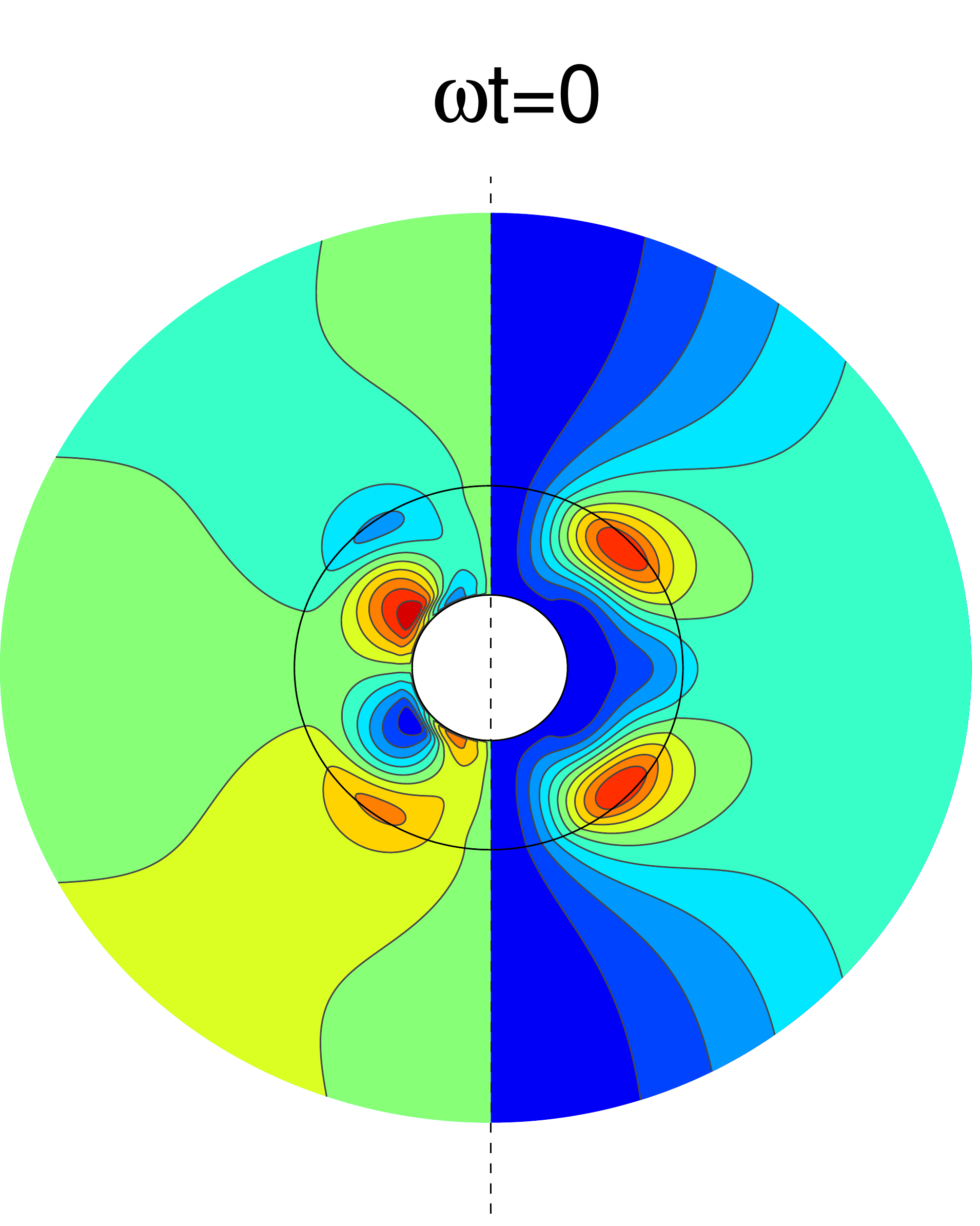}
\caption{
Same as model J
but with a convection zone which extends from $x=0.5$. In this case $\calpha=7.86$, $\clambda=49.5$ and $\rat=0.90$. 
\label{fig2}}
\end{center}
\end{figure}
\begin{figure}
\begin{center}
 \includegraphics[width=0.32\textwidth, angle=-90]{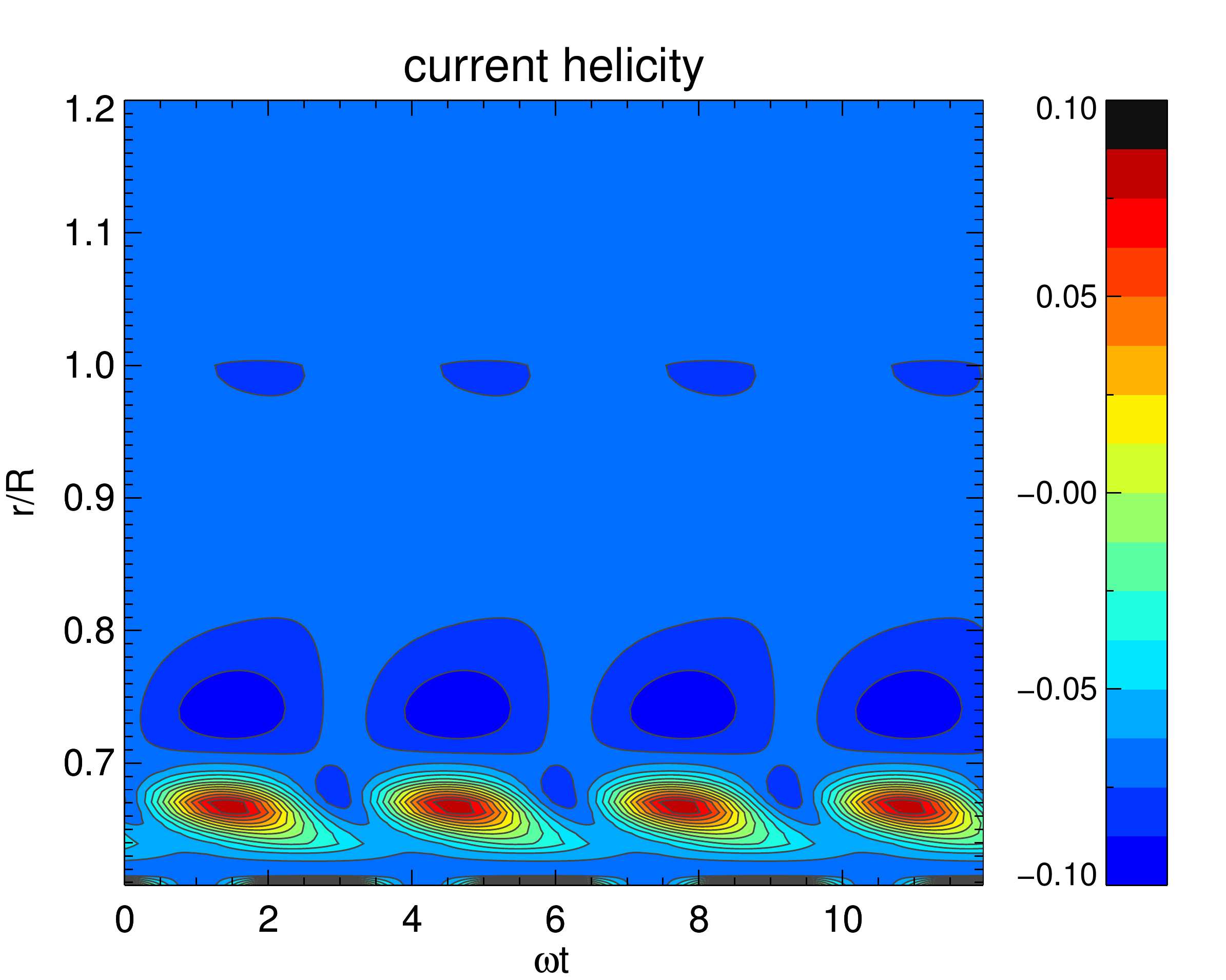}
 \includegraphics[width=0.32\textwidth, angle=-90]{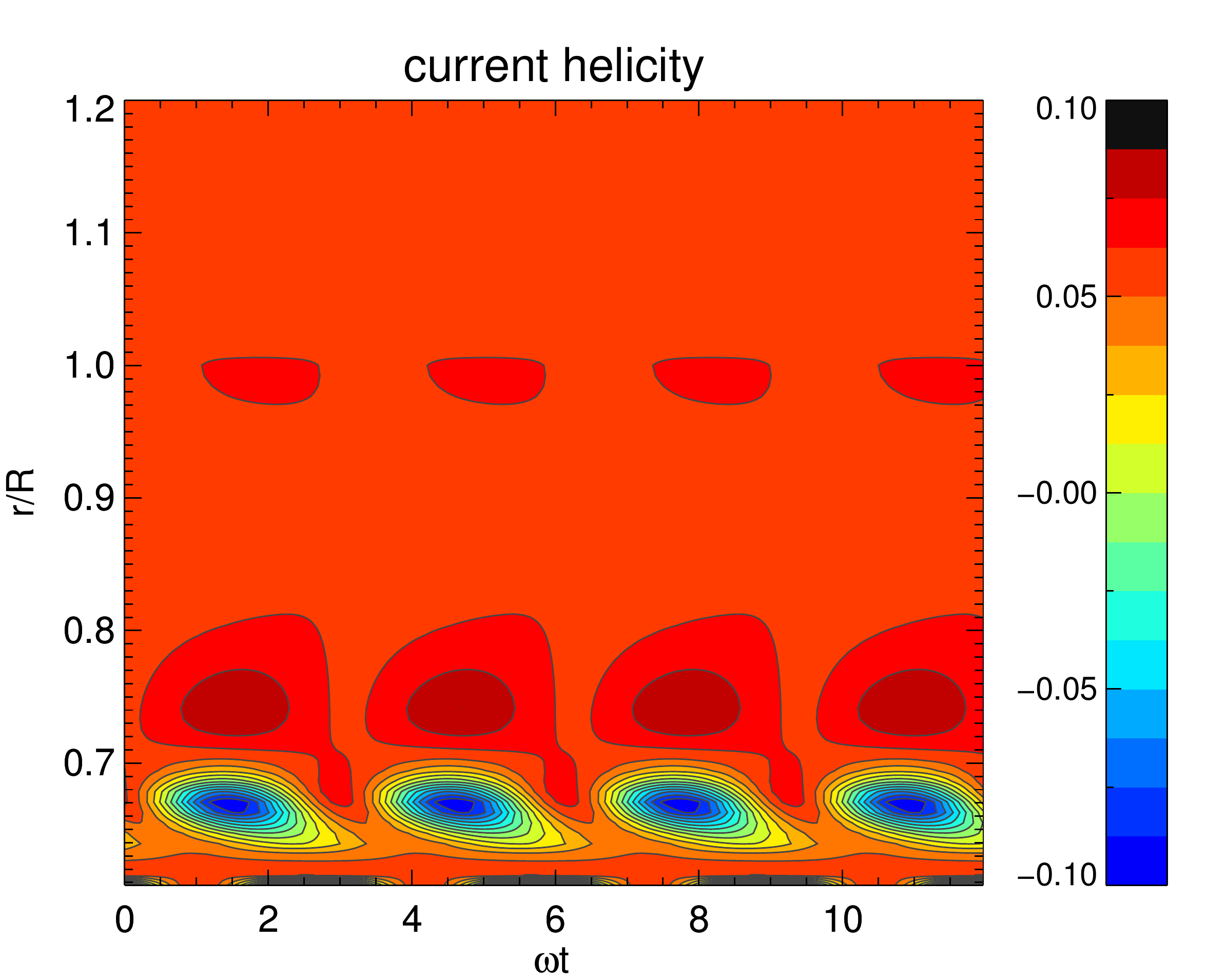}
\caption{
Upper panel: evolution of  $\mu_0 R \mathbf{J}\cdot\mathbf{B}/\langle \mathbf{B}^2 \rangle$  computed at +30$^\circ$ latitude
for a $\alpha^2 \Omega$ advection dominated solar dynamo model with 
$k R=0.1$, $\calpha=3.03$, $\comega=3 \cdot 10^4$, $\cur=400$ and the cycle period is 32 years. 
Lower panel: same model as upper panel but opposite hemisphere. Notice that the helicity changes sign from the 
turbulent zone in the interior, to the exterior.
\label{fig3}}
\end{center}
\end{figure}
\section{Dynamo models}
Let us specialize our formalism to the case of a solar-like star. We  assume the following form for the velocity field
$\mathbf{U}=\mathbf{u}(r,\theta)+ r \sin\theta \Omega(r,\theta) \mathbf{e}_\phi$ and
we model the solar-like differential rotation in the following way:
\be
\Omega(r,\theta) =\Omega_c +\frac{1}{2}\left [ 1+{\rm erf}\left (\frac{x-x_c}{d}\right)\right](\Omega_s(\theta)-\Omega_c)
\ee
where  $x_c$ is the location of the convection zone in units of the stellar radius,  
$d=0.02$, $\Omega_c$ is the uniform angular velocity of the radiative core,
$\Omega_s(\theta)=\Omega_0- d\Omega\cos^2\theta$. Here   $d\Omega=\Omega_{eq}-\Omega_c$ 
is the surface differential rotation (in principle obtained from
observations).  For actual calculations we fixed  $x_c=0.7$ and $\Omega_c/\Omega_{eq}= 0.9$ but 
$\dome$ is allowed to vary.  The radial profile of the turbulent diffusivity is assumed to be the following: 
\be\label{eta}
\eta=\eta_c +\frac{1}{2}(\eta_t-\eta_c)\left [ 1+{\rm erf}\left(\frac{x-x_x}{d}\right)\right ]
\ee
with $\eta_c/\eta_t=10^{-1}$.  
The $\alpha$ effect is proportional to $\cos\vartheta$ and its radial profile is assumed to be 
uniformly distributed in all the convection zone (see \cite{bonanno02} for details).
The meridional circulation $\mathbf{u}$ is obtained from the stream function $S(r)$
which as explained in \cite{bonanno13}  $S(r)$ can be obtained from an underlying stellar model. 
The flow as usual is equatorwards at the equator and poleward at the surface.

As usual, let us introduce the following dynamo numbers, 
$\comega ={R^2\Omega_0}/{\eta_t}$,
$\calpha={R\alpha_0}/{\eta_t}$, 
$\cur={R U_0}/{\eta_t}$,
$\clambda={R^2\omega}/{\eta_t}$
where $\Omega_0$ is the rotation rate at the equator and $U_0$ is the maximumm strength of $u_\theta$ 
at the bottom of the convection zone.
The resulting eigenvalue problem  can be conveniently solved by inverting a block-diagonal 
complex matrix \citep{rad73,bonanno13} which must be truncated to the desired 
numerical accuracy.  For calculations we used the kinematic dynamo  code CTDYN, 
developed by the author and extensively tested in \cite{jouve10}. 
Our  results are summarized in table (\ref{tab1}). 

\begin{table}[h]
\begin{center}
\begin{tabular}{cccccccc}
${\cal M}$ & $\comega$ & $\cur$ & $\dome$ & $k R$ & $\calpha$ & $\clambda$ & $\rat$ \\
\hline
A &1000 & 0  &0.1 & 0.1 & 7.87 & $\infty$ & 0.67 \\
B &1000 & 0  &0.1 & 0.4 & 7.88 & $\infty$ & 0.67 \\
C &1000 & 0  &0.3 & 0.1 & 11.24& $\infty$ & 0.67 \\
D &1000 & 100 &0.3 & 0.1 & 7.78& $\infty$ & 0.37 \\
E &2000 & 0  &0.1 & 0.1 & 9.49 & $\infty$ & 0.63 \\
F &2000 & 0  &0.3 & 0.1 & 10.97 & 79.02 & 0.91 \\
G &2000 & 0  &0.3 & vac & 11.94 & 69.98 & 0 \\
J &4000 & 0  &0.1 & 0.1 & 10.11 & 62.93 & 0.85 \\
K &4000 & 0  &0.1 & 0.5 & 10.04 & 63.03 & 0.86 \\
L &4000 & 100 &0.1 & 0.1 & 5.57 & $\infty$ & 0.68 \\
M &4000 & 200 &0.1 & 0.1 & 5.13 & $\infty$ & 0.37 \\
\end{tabular}
\caption{Summary of the numerical simulations. In particular it shows how $\rat$
depends on $\comega$ and various other input parameters (``vac" stands for current-free boundary condition). 
${\cal M}$ is  the model name, and in particular 
for model J the toroidal and poloidal field are displayed  in \fig{fig1} at various values of the cycle phase.  \label{tab1}}
\end{center}
\end{table}

Let us first stress 
that the $\rat$ at the surface is non-zero and it is now an increasing function of $\comega$, an effect
clearly expected for an $\alpha\Omega$ dynamo.
The surface differential rotation also plays an important role because it strongly influences
the value of $\rat$, as it can be deduced by looking at models E and F for instance. 
On the contrary, it should also be noticed the decrease 
of $\rat$  as the flow is increased (see models J, L, M for instance). 
This fact has a clear physical interpretation: as the flow is poleward below the surface, the 
toroidal belts become more and more confined below the surface at high $\cur$.
It is also reassuring to notice that, as long as  $  k R\ll1 $ our results are not
sensitive to the choice of $\xi$. 
On the other hand, with the new boundary 
conditions the value of the critical $\calpha$ is in general 
smaller  than with the standard   current-free boundary conditions
as it can be observed by 
looking at models F and G. We also verified that our results are not qualitatively 
dependent on the location of the external boundary $\ro$.  
Moreover, low mass stars with more extended convection zone have in general higher values
of $\rat$, if $\comega$ is large enough, as discussed in the model of \fig{fig2}.

 If the dynamo action is instead driven by the meridional circulation (as in mean field models of the solar dynamo)
the new boundary condition does not alter the internal dynamo action because the toroidal field is localized
at the bottom of the convection zone. 

{In the case of a generic $\alpha\Omega$ advection-dominated 
dynamo action we observe a change of sign of the dimensionless ratio
$\mathbf{J}\cdot\mathbf{B}/\langle \mathbf{B}^2 \rangle$  
from positive  (in the norther hemisphere) in the turbulent zone in the interior to negative
in the exterior. The opposite happens in the southern hemisphere as one can see in \fig{fig3}. 
A similar change of sign between the turbulent zone and the exterior has also been observed
in \citep{warne11}, although in our mean-field models the current helicity is mostly negative 
in the outer layers in the norther hemisphere.}

\section{conclusions}
The description of the external field in terms of solution of the Helmholtz equation allowed us to extrapolate the internal toroidal 
field generated by the dynamo on the photosphere, and finally to make contact with the observations. The assumption beyond this idea 
is the possibility of treating the corona as an external passive medium with effective macroscopic dielectric properties, if averaged
over long enough time scales. Although this approach oversimplifies the complex physics of the corona, in our opinion 
it represents a significant improvement
of the  current-free  boundary conditions for which $\BT\equiv0$ on the surface, at least in some class of very active stars. 
The resulting dynamo numbers are 
in general smaller than the standard critical dynamo numbers; the ratio $\rat$ increases with $\comega$
and with a more
extended convection zone, and decreases with $\cur$. 
Indeed, fast rotating stars with larger convection zone should approach  a cylindrical rotation law in the interior
with a smaller surface meridional circulation.

One can therefore argue that the general increase of the surface toroidal energy in low mass fast rotating
stars finds its natural explanation in an underlying $\alpha\Omega$ dynamo mechanism. 

A detailed study of all the parameter space and a comparison with the global topologies inferred from observations
will be discussed in a longer paper. We also plan to extend this investigation including non-axisymmetric solutions
of higher azimuthal modes.  

\acknowledgments
I would like to thank the colleagues of the MHD group of the Leibniz Institute for Astrophysics in Potsdam for important
comments and hospitality. I am also indebted to Rim Fares for clarifications 
on Zeeman-Doppler imaging and on magnetic field topology reconstruction and to the anonymous 
referee for his constructive criticism.

\end{document}